\documentclass[pre,superscriptaddress]{revtex4}
\usepackage{amsmath}
\usepackage{graphicx}

\newcommand{\fraction}{\lambda}
\newcommand{\cheat}{C}

\begin{document}

\title{Pattern formation during diffusion limited transformations in solids}

\author{M. Fleck} 
\affiliation{Institut f{\"u}r Festk{\"o}rperforschung, Forschungszentrum J{\"u}lich, D-52425 J{\"u}lich, Germany}
\author{C. H{\"u}ter} 
\affiliation{Institut f{\"u}r Festk{\"o}rperforschung, Forschungszentrum J{\"u}lich, D-52425 J{\"u}lich, Germany}
\author{D. Pilipenko} 
\affiliation{Institut f{\"u}r Festk{\"o}rperforschung, Forschungszentrum J{\"u}lich, D-52425 J{\"u}lich, Germany}
\author{R. Spatschek}
\affiliation{Interdisciplinary Centre for Advanced Materials Simulation (ICAMS), Ruhr-Universit{\"a}t Bochum, Germany}
\affiliation{Institut f{\"u}r Festk{\"o}rperforschung, Forschungszentrum J{\"u}lich, D-52425 J{\"u}lich, Germany}
\author{E.~A. Brener}
\affiliation{Institut f{\"u}r Festk{\"o}rperforschung, Forschungszentrum J{\"u}lich, D-52425 J{\"u}lich, Germany}

\date{\today}

\begin{abstract}
We develop a description of diffusion limited growth in solid-solid transformations, which are strongly influenced by elastic effects.
Density differences and structural transformations provoke stresses at interfaces, which affect the phase equilibrium conditions.
We formulate equations for the interface kinetics similar to dendritic growth and study the growth of a stable phase from a metastable solid in both a channel geometry and in free space.
We perform sharp interface calculations based on Green's function methods and phase field simulations, supplemented by analytical investigations.
For pure dilatational transformations we find a single growing finger with symmetry breaking at higher driving forces, whereas for shear transformations the emergence of twin structures can be favorable.
We predict the steady state shapes and propagation velocities, which can be higher than in conventional dendritic growth.
\end{abstract}

\maketitle

\section{Introduction}

Transformations between different solid states of a material are essential for many technological and scientific applications, and understanding their kinetics is therefore not only interesting because of the scientific variety and beauty, but also important for tailoring of new materials with specific properties.
Often, these processes are accompanied by elastic deformations, which can be e.g.~due to density differences or structural changes, which provoke stresses at interfaces between adjacent phases.
The influence on the thermodynamics of the transitions between different phases has been thoroughly discussed in the literature \cite{khachaturyan, Roitburd1974}, whereas the kinetics of these processes are still far less understood.

Facing this lack of understanding, we would like to mention that very fruitful concepts have been developed to describe the dynamics of solidification or melting processes as moving boundary problems.
It means that the evolution of the different phases is described on the very detailed level of the motion of the interfaces between them, for which suitable equations of motion have to be posed.
A particular application is the famous example of dendritic growth, where heat or component diffusion is the rate limiting process.
At the interface between the solid and melt phase the temperature or concentration is close to its phase equilibrium value.
The concentration jump or the release of latent heat due to partitioning leads to a jump in the corresponding fluxes at the interfaces, and the magnitude of this jump is directly related to the local interface velocity.
Altogether, this provides a closed description for the motion of each interface point, and many efforts have been done towards analytical and numerical solutions of this problem.

It turns out that the questions concerning existence, shape and growth velocity of steady state patterns in solidification crucially depend on selection mechanisms.
In this spirit the influence of many different physical effects on solidification or melting processes has been studied throughout the years \cite{kessler, brenermelnikov}.
For example, the effect of isotropic surface tension was proven not to serve as a selection mechanism for a solution, whereas anisotropic surface tension leads to a unique solution. 
Also, the appearance of triple junctions can provide a selection mechanism, and can therefore lead to a solution of the free boundary problem \cite{BrenerMeDenisTemkin2007}.

Of course, the situation considering solid-solid transformation kinetics differs significantly from the solidification or melting problem, where elastic effects often play only a minor role.
Still, we follow here the same successful concepts to formulate a free boundary description for the transformation kinetics in solids, now including the important effect of elasticity.
In real systems many other effects apart from elastic effects come into play -- like inhomogeneous compositions, the Mullins-Sekerka instability, crystal anisotropies, polycrystalline structures, etc. -- and also quite successful attempts have been made to model this complex behavior (see for example in \cite{SteinbachApel} and references therein).
Here, in contrast, we focus on gaining a fundamental understanding of the elastic influence on pattern formation processes, and we restrict our investigations therefore to simple and ``clean" model systems. 
Correspondingly, we consider only a single diffusion field, namely the thermal diffusion in pure materials.

The formulation of such a model system requires to write down explicit expressions for the boundary conditions and suitable equations of motion at the propagating interfaces, which therefore have to be tracked during the entire dynamical process.
Finally, it provides a detailed description of the microstructure evolution on the level of individual grains in a polycrystalline structure.
As for the dendritic growth, we focus here on the regime of diffusion limited growth.
Then, propagation of a front generates or consumes latent heat, which has to diffuse through the system, and we assume that this is the rate-limiting process.
This is in contrast to isothermal solid-solid transformations that were discussed in \cite{BrenerMarchenkoSpatschek2007}, which are driven by local interface kinetics, therefore allowing fast front propagation also on the scale of the sound speed. 
Many important properties result from the influence of elastic effects due to structural transformations and density differences.

Apart from sharp interface methods, that are based on Green's function techniques, we also use phase field descriptions, which are nowadays widely used.
More importantly, however, is that we supplement these numerical investigations by analytical calculations which allow to understand these processes and the relevant parameters on a deeper basis.

\section{Formulation of the Problem}

\begin{figure}
\begin{centering}
\includegraphics[width=0.55\textwidth]{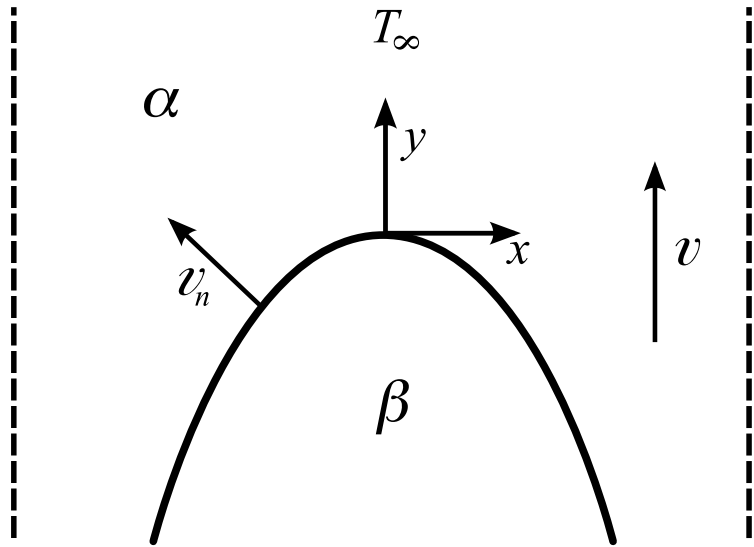}
\par\end{centering}
\caption{
Schematic figure of the steady state growth, with constant velocity $v$, of a thermodynamically favored phase $\beta$ into a metastable phase $\alpha$, which is initially at temperature $T_{\infty}$.
The motion of the interface is locally described by the interface-normal velocity $v_{n}$, which is given by the local heat conservation Eq.~(\ref{eq:Heat-conservation-equation}).
The dashed lines at the two side borders indicate that we will discuss two different types of outer boundary conditions:
Free growth, where the strains decay for $x,y\rightarrow \infty$, and channel growth, with fixed displacements and thermal insulation at the walls.
In the growth direction $y$ the system is assumed to be infinite in both cases.
\label{fig:Geometry}}
\end{figure}

Specifically, we consider the growth of an equilibrium solid phase $\beta$ from another metastable solid phase $\alpha$,
as depicted in Fig.~\ref{fig:Geometry}.
At a temperature $T_{eq}$, both stress-free phases are in bulk equilibrium with each other, whereas below this value the new $\beta$ phase is thermodynamically favorable.
Far ahead of the propagating front, the system is therefore exposed to a lower temperature $T_\infty$, and we define a dimensionless temperature $w=\cheat(T-T_{\infty})/L$, where $L$ denotes the latent heat of the transition and $\cheat$ the heat capacity.
Within the solid phases the evolution of the temperature field obeys a diffusion equation,
\begin{align}
D\nabla^{2}w= & \frac{\partial w}{\partial t}, \label{eq:Diffusion-dimless-Temperature}
\end{align}
where $D$ is the thermal diffusivity, which we assume to be equal in both phases, as well as the heat capacity (symmetrical model).

The motion of the interfaces follows from energy conservation principles, which account here for the net heat flux and the generation or absorption of latent heat due to the local motion of the interface with normal velocity $v_n$,
\begin{align}
v_{n}= & D\mathbf{n}\cdot(\nabla w_{int}^{[\beta]}-\nabla w_{int}^{[\alpha]}), \label{eq:Heat-conservation-equation}
\end{align}
where $\bf n$ is the interface normal and the superscripts denote the phases.

Elastic effects enter into the problem via a modification of the local phase equilibrium condition at the moving interfaces.
In contrast to previous investigations \cite{BrenerIordanskiiMarchenko1999, BrenerMarchenko1992}, we now assume the phases to be coherently connected, i.e.~we neglect the appearance of any slips, detachments or defect formation at the interfaces, which is reasonable for small misfits. 
It is important to mention here, that with the assumption of coherency we formulate a completely different model being capable to describe qualitatively different types of elastic effects.
For example the well-known effect of elastic hysteresis \cite{khachaturyan, Roitburd1974}, which was proven to disappear without coherency \cite{BrenerMarchenko1992}, is now included in the model.

In a Lagrangian description of linear elasticity, the displacement field is denoted by $\mathbf{u}$, and the coherency condition reads, 
$\mathbf{u}^{[\alpha]}=\mathbf{u}^{[\beta]}$ at the interface.
The strain field is then given by the symmetrical spatial derivative, $\epsilon_{ik}=\left(\partial u_{i}/\partial x_{k}+\partial u_{k}/\partial x_{i}\right)/2$.
This implies that tangential and shear strains $\epsilon_{\tau\tau},\epsilon_{s\tau},\epsilon_{ss}$ are continuous at the interface with $\tau,s$ being the two tangential directions. 
Force balance demands the continuity of the normal and shear stresses, $\sigma_{nn},\sigma_{n\tau},\sigma_{ns}$, where $n$ is the normal direction.
Here, the stresses are defined as the derivative of the free energy with respect to the strains, $\sigma_{ik}=\partial F/\partial\epsilon_{ik}$.

The interface temperature for a planar, stress free interface is given by $w=\Delta$, with $\Delta=\cheat(T_{eq}-T_{\infty})/L$.
Capillary corrections induce a curvature dependent term, and for simplicity we discuss only the case of isotropic surface tension $\gamma$.
Finally, elastic deformations lead to an additional change of the interface temperature. 
Altogether, this leads to the local equilibrium boundary condition \cite{DenisBrenerMe2008} 
\begin{align}
w_{int} & =\Delta-d_{0}\kappa+\frac{T_{eq}\cheat}{L^{2}}\delta \tilde{F}_{el},\label{eq:Phase-equilibrium-condition}
\end{align}
with $d_{0}=\gamma T_{eq}\cheat/L^{2}$ being the capillary length and  $\kappa$ the local curvature of the interface, counted positive for a convex phase $\beta$ (see Fig.~\ref{fig:Geometry});
Notice that due to the coherency constraint \cite{Privorotskii1971, SpatschekFleck2007}, the elastic shift of the equilibrium temperature is proportional to the difference of a new potential $\delta \tilde{F}_{el} =\tilde{F}_{el}^{[\alpha]} - \tilde{F}_{el}^{[\beta]}$, which is related to the elastic free energy $F_{el}$ via the Legendre transformation $\tilde{F}_{el} = F_{el} -\sigma_{ni}\epsilon_{ni}$.
Here, we use the sum convention for repeated indices for the sake of brevity.

From now on, we restrict our considerations to linear isotropic elasticity.
Taking the relaxed state of the $\alpha$ phase as reference, the elastic contribution to the free energy there reads
\begin{align}
F_{el}^{[\alpha]} & =\frac{E}{2(1+\nu)}\left(\frac{\nu}{1-2\nu}\epsilon_{ii}^{2}+\epsilon_{ik}^{2}\right),\label{eq:Free-energy-reference}
\end{align}
where $E$ and $\nu$ are Young's elastic modulus and Poisson ratio, respectively. 
In contrast, the elastic contribution to the free energy density of the new phase $\beta$ reads
\begin{align}
F_{el}^{[\beta]} & =\frac{E}{2(1+\nu)}\left(\frac{\nu}{1-2\nu}\left(\epsilon_{ii}-\epsilon_{ii}^{0}\right)^{2}+\left(\epsilon_{ik}-\epsilon_{ik}^{0}\right)^{2}\right),\label{eq:Free-energy-new}
\end{align}
which has an other state of zero elastic energy due to the lattice strain $\epsilon_{ik}^{0}$ assigned to the phase transformation (see for example \citet{khachaturyan, Roitburd1974}).
In the following we will call $\epsilon_{ik}^{0}$ the eigenstrain of the transformation. 
Particular cases of eigenstrains and their interpretation will be discussed in the next section.

The mechanical equilibrium and coherency conditions provide expressions for the discontinuous jumps of the strains at the interface
\begin{align}
\epsilon_{nn}^{[\beta]}-\epsilon_{nn}^{[\alpha]}= & \epsilon_{nn}^{0}+\frac{\nu}{1-\nu}(\epsilon_{ss}^{0}+\epsilon_{\tau\tau}^{0}),\label{eq:Strain-Jump-n}\\
\epsilon_{n\tau}^{[\beta]}-\epsilon_{n\tau}^{[\alpha]}= & \epsilon_{n\tau}^{0},\quad\epsilon_{ns}^{[\beta]}-\epsilon_{ns}^{[\alpha]}=\epsilon_{ns}^{0},\label{eq:Strain-Jump-tau/s}
\end{align}
which are only related to the imposed eigenstrain. Defining an eigenstress tensor via Hooke's law 
\begin{align}
\sigma_{ik}^{0} & =\frac{E}{1+\nu}\left(\epsilon_{ik}^{0}+\frac{\nu}{1-2\nu}\delta_{ik}\epsilon_{ll}^{0}\right),\label{eq:Eigenstress-definition}
\end{align}
we obtain after a few straightforward algebraic manipulations for the elastic contribution to the local equilibrium condition
Eq.~(\ref{eq:Phase-equilibrium-condition})
\begin{align}
\delta\tilde{F}_{el}= & \sigma_{ik}^{0}\epsilon_{ik}^{[\alpha]} - \frac{E}{2(1-\nu^{2})}\left((\epsilon_{\tau\tau}^{0})^{2}+(\epsilon_{ss}^{0})^{2}+2\nu\epsilon_{\tau\tau}^{0}\epsilon_{ss}^{0}+2(1-\nu)(\epsilon_{\tau s}^{0})^2\right).
\label{eq:Phase-equilibrium-elast-cont}
\end{align} 
Apparently, this expression depends on the strain state of the $\alpha$ phase at the interface.

As already implicitly assumed above, the elastic degrees of freedom relax fast on diffusive timescales, and therefore the application of static elasticity is legitimate.
Therefore, Newton's second law becomes
\begin{align}
\frac{\partial\sigma_{ik}}{\partial x_{k}} & =0.\label{eq:ElastoStatics}
\end{align}

The above set of equations now has to be supplemented by boundary conditions at the external system boundaries, and we will discuss different scenarios below.
They complete the self-sonsistent description of the moving boundary problem, which require the simultaneous solution of the elastic and diffusion equations, coupled via the boundary conditions at the propagating interfaces, leading to a complicated nonlinear and nonlocal problem.

We note that we will later also discuss more complicated situations with more than one growing phase.
Then, also similar boundary conditions and equations have to be set up for the interfaces between them.
Typically, then also triple junctions will appear, where the contact angles are given by Young's law.


\section{Models of structural transitions}

The simplest case of a characteristic lattice strain is to assume the bond length of the new phase $\beta$ to be uniformly longer or shorter in all directions in comparison to the reference phase, i.e.~$\epsilon_{ik}^{0}=\epsilon_{ik}^{d}=\epsilon\delta_{ik}$, which implies a density change.
Notice, that this kind of transformations, to which we will refer to as {\em dilatational eigenstrain}, only changes the volume of an elementary cell, and not its relative dimensions or its scaled shape.

\begin{figure}
\begin{centering}
\includegraphics[width=0.6\textwidth]{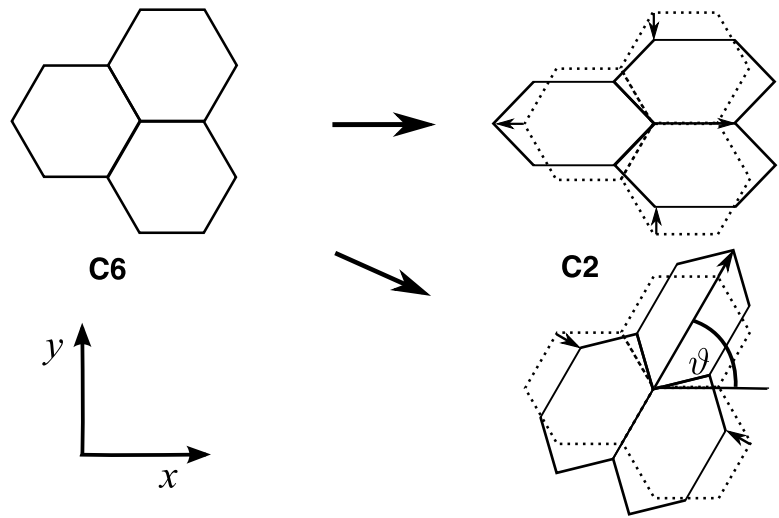} 
\par\end{centering}
\caption{\label{fig:HexagonalToOthorombic}
Visualisation of the hexagonal to orthorhombic transformation.
From the symmetry its obvious, that  the orientation angle $\vartheta$ of the new orthorhombic phase can have only the three different values $0,\pm2\pi/3$.}
\end{figure}

The opposite case are transformations, where the volume of the unit cell is conserved and only the shape is changed.
These {\em shear transformations} are characterized by a traceless eigenstrain tensor.
A particular transition involving shear strain can occur e.g.~in hexagonal crystals, where the transition leads to a lowering of the symmetry from $C_{6}$ to $C_{2}$. 
This is the case for example in hexagonal-orthorombic transitions in ferroelastics (see \cite{CurnoeJacobs} and references therein). 
Let the principal axis $C_6$ be oriented in $z$ direction.
For simplicity we neglect all other possible strains with higher (axial) symmetry.
By proper choice of the crystal orientation around the main axis in the initial phase, we obtain the new phase in three possible states in the $xy$ plane due to the original hexagonal symmetry (See Fig.~\ref{fig:HexagonalToOthorombic}). 
Then the nonvanishing components of the eigenstrain tensor $\epsilon_{ik}^0 = \epsilon_{ik}^{s}$ are $\epsilon_{yy}^{s}=-\epsilon_{xx}^{s}=\epsilon\cos2\vartheta$ and $\epsilon_{xy}^{s}=\epsilon\sin2\vartheta$, where the orientation
angle $\vartheta$ is either $\vartheta=0$ or $\vartheta=\pm2\pi/3$.


We also consider a more generic mixed mode case, where the self-strain of the transformation is a linear combination of the above two different transformations. 
We will discuss the following two mixing cases
\begin{align}
\epsilon_{ik}^{0}=\epsilon_{ik}^{\pm}(\eta) &= \eta\epsilon_{ik}^{d} \pm (1-\eta)\epsilon_{ik}^{s}
\label{MixedModeEigenstrainDefinition}
\end{align}
and refer to them as positive or negative mixing.

It is important to mention that switching from positive to negative mixing corresponds to a rotation of the eigenstrain tensor by an angle 
$\pi/2$.  
Since our model is isotropic in all respects except the choice of the eigenstrain tensor, the only predefined direction is given by the growth velocity. 
Therefore, we can interpret the negative mixing case to represent physically the same system as the positive mixing case but with a growth direction perpendicular to it.



\section{Free growth}

As a first application, we discuss the growth of the stable $\beta$ phase from the metastable $\alpha$ phase in an infinitely large system, which is exposed to the temperature $T_\infty$ far away from the interface.
For simplicity, we discuss an effectively two-dimensional infinite system, assuming translational invariance in $z$ direction.
In particular, we assume a plane strain situation from point of view of elasticity.
Also, we choose the elastic boundary conditions such that all stresses decay far away from the interface.


Seeking for steady state solutions in free space, we use Green's function methods for solving both the elastic and the diffusional problem, to finally obtain self-consistently the shape and the velocity of the growing front.
While the derivation of the integral representation of the thermal field is well-known, see e.g.~\cite{Langer1980}, we present the calculation of $\delta\tilde{F}_{el}$ in more detail \cite{DenisBrenerMe2008}.
To this end, we express the elastic problem using the eigenstrain-independent stress tensor $\tilde{\sigma}_{ik}$;
for the $\alpha$ phase it is the same as the nomimal stress, $\tilde{\sigma}_{ik}^{[\alpha]}=\sigma_{ik}^{[\alpha]}$, whereas in the $\beta$ phase it is $\tilde{\sigma}_{ik}^{[\beta]}=\sigma_{ik}^{[\beta]}+\sigma_{ik}^{0}$.
In terms of the new stress tensor $\tilde{\sigma}_{ik}$, the mechanical equilibrium equation $\partial\tilde{\sigma}_{ik}/\partial x_{k}=f_{i}$ introduces a force density  which is localized on the interface and vanishes in the bulk,
\begin{eqnarray}
f_{i} & = & \tilde{\sigma}_{in}^{[\beta]}-\tilde{\sigma}_{in}^{[\alpha]}=\sigma_{in}^{0}. \label{eq:Artificial-Interface-Forces}
\end{eqnarray}
This means that the problem of two coherently connected solids with different reference state is equivalent to a monolithic material with a distribution of point forces along the interface.
We can therefore use an integral representation of the displacement field,
 %
%
\begin{eqnarray}
u_{i} & = & \int ds^{\prime}G_{ik}({\bf r}-{\bf r}^{\prime})f_{k}({\bf r}^{\prime}),\label{formal_solution_displacements}
\end{eqnarray}
where the integration is performed along the interface, and the free-space Green's tensor $G_{ik}({\bf r}-{\bf r}^{\prime})$ of the elastic
problem is given by \cite{LandauLifshitzElasticity}
\begin{equation}
G_{ik}({\bf r}) = \frac{1+\nu}{4\pi(1-\nu) E} \left( \frac{x_i x_k}{r^2} - (3-4\nu)\delta_{ik} \ln r \right).
\end{equation}
This function is the displacement response at ${\bf r}$ to a point force located in the origin.
As a result, the mismatch provokes long-ranged elastic deformations, and therefore induces a nonlocal character to the problem.
The above expressions can then be used to calculate the elastic influence on the interface temperature $w_{int}$.


Far behind the tip the interface profile becomes parabolic and is described by the asymptotic Ivantsov parabola \cite{Ivantsov}.
In this region, the elastic strains have decayed, and only the constant local contribution to $\delta \tilde{F}_{el}$ (second term in Eq.~\ref{eq:Phase-equilibrium-elast-cont}) remains.
Consequently, the effective driving force $\tilde{\Delta}$ consists not only of the thermal undercooling $\Delta$, which is controlled by the applied temperature $T_\infty$, but also a contribution $\Delta_{el}$ from the elasticity,
\begin{align}
\Delta_{el}= &  \frac{\cheat T_{eq}E [(\epsilon_{yy}^{0})^{2} + (\epsilon_{zz}^{0})^{2} + 2\nu \epsilon_{yy}^0 \epsilon_{zz}^0]}{2(1-\nu^{2})L^2}.
\label{eq:Definition-DeltaEL}
\end{align}
We can consider $\Delta_{el}$ as a parameter quantifying the strength of the elastic effects. 
The elastic effects therefore induce a shift of the equilibrium temperature, and the effective undercooling $\tilde{\Delta}$ is lower than the actually applied thermal driving force,
\begin{align}
\tilde{\Delta} & =\Delta-\Delta_{el}, \label{eq:Shifted-undercooling}
\end{align}
which reflects the elastic hysteresis \cite{khachaturyan, Roitburd1974}.

By elimination of the thermal field \cite{Langer1980} we can derive the steady state equation for the shape of the solid-solid interface in a closed dimensionless representation, which reads in the co-moving frame of reference \cite{DenisBrenerMe2008}, 
\begin{eqnarray}
\Delta-\frac{d_{0}\kappa}{R}+\frac{T_{eq}\cheat\delta\tilde{F}_{el}}{L^{2}} & = & \frac{p}{\pi}\int_{-\infty}^{\infty}dx'\exp[-p(y(x)-y(x'))]K_{0}(p\mu(x,x')).
\label{MainPartClosedEquation}
\end{eqnarray}
Here, $p=vR/2D$ denotes the Peclet number ($v$ is the steady state velocity), and it is related to the driving force by the implicit relation $\tilde{\Delta}=\sqrt{\pi p}\,e^{p}\,\mathrm{erfc}(p)$.
Furthermore, $K_0$ is the modified Bessel function of third kind in zeroth order, and $\mu(x,x')=[(x-x')^{2}+(y(x)-y(x'))^{2}]^{1/2}$.
In the above expression we rescaled all lengths by the radius of curvature $R$ of the asymptotic Ivantsov parabola. 


Consequently, the central equation (\ref{MainPartClosedEquation}), together with the expression for the elastic contribution (\ref{eq:Phase-equilibrium-elast-cont}), where the strains are expressed via the integral representation (\ref{formal_solution_displacements}) results in a complicated integro-differential equation for the interface shape.
It differs from conventional dendritic growth by the appearance of the elastic term.
To better understand its influence, we briefly review the solvability conditions for classical dendritic growth \cite{kessler,brenermelnikov}, for situations with pure dilatational and shear strain \cite{DenisBrenerMe2008}, and cases with a triple junction \cite{BrenerMeDenisTemkin2007}:
In classical dendritic growth, any solution for isotropic surface tension with nonzero solvability parameter $\sigma=d_{0}/pR$ must exhibit a finite negative tip kink angle $\phi$. 
For this reason, dendritic growth is not possible for isotropic surface energy, since no smooth dendrite profiles exist.
A dilatational eigenstrain simply shifts the equilibrium interface temperature, and therefore suffers from the same missing selection as the dendrite problem without surface tension anisotropy.
The presence of a shear eigenstrain, which will also be discussed in more detail below, leads to the propagation of a bicrystal.
In this case, the elastic effects operate as new selection mechanism.
Independent of this elastically induced selection mechanisms, the presence of a triple junction alone also works as a selection mechanism.

We solve the eigenvalue problem Eq.~(\ref{MainPartClosedEquation}) numerically for the steady state shape and the stability parameter $\sigma=\sigma^*$ in the spirit of Ref.~\cite{meiron}, and the results are presented in Section \ref{results}.

\section{Channel Growth}

The growth of the $\beta$ phase in a channel of finite width $W$ differs significantly from the free growth scenario discussed above.
The specific behavior depends of course on the boundary conditions, and we assume fixed grip conditions on the long sides of the channel, i.e. the displacement is constant.
\begin{figure}
\begin{center}
\includegraphics[width=0.65\textwidth]{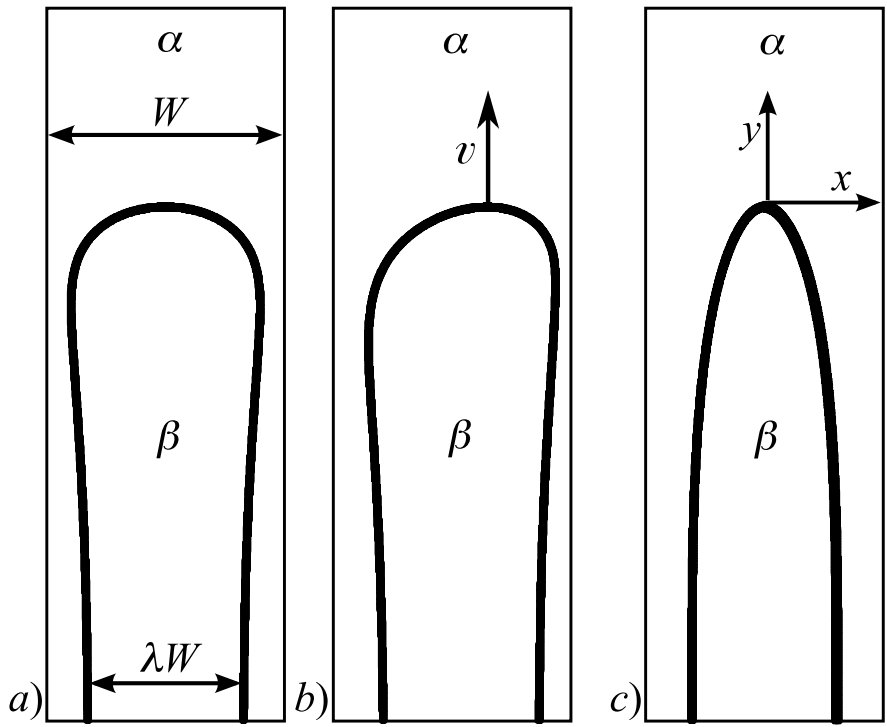}
\caption{Phase field shapes of single $\beta$ phases growing with constant velocity $v$ into the metastable $\alpha$ phase in a finite channel of width $W$. 
The shapes in $a)$ and $b)$ correspond to the dilatational case ($\eta = 1$), with $\Delta_{el}=0.4$ and an undercooling, which is chosen such that according to Eq.~(\ref{eq:Fraction-of-beta-phase}) it would correspond to an asymptotic phase fraction $\fraction=0.65$.
While $a)$ shows the shape of the symmetrical finger, $b)$ shows the shape of the sightly faster parity broken finger (see Fig.~\ref{fig:DiagonalEigenstrain}a).
In panel $c)$ we have plotted the shape in the case of negative mixing eigenstrain, $\eta=0.5$, which in that case corresponds to growth in the privileged direction.
The parameters are $\Delta_{el}=0.05$ and $\fraction=0.6$ (for the resulting velocity see Fig.~\ref{fig:DiagonalEigenstrain}b). }
\label{DilatationalSetup}
\end{center}
\end{figure}
Again, we discuss an effectively two-dimensional system by the assumption of translational invariance in $z$ direction (see Fig.~\ref{DilatationalSetup} for the setup).
Furthermore, we assume that the $\alpha$ phase far in front of the growing tip is fully relaxed, which is in agreement with the boundary conditions $u_x\equiv 0$ and $u_y\equiv 0$ along the boundaries of the whole, infinitely long strip.
From a thermal point of view, we assume thermal insulation along these boundaries, so no temperature flux through the surfaces, $\partial T/\partial x=0$ on the outer boundaries.
Far ahead of the tip, the temperature is given by the dimensionless undercooling $\Delta$.

Due to the phase transition, the release of the latent heat leads to a heating of the system, therefore the temperature far behind the tip $T_{-\infty}$ is higher than in front.
Under the given circumstances it is quite obvious that in this tail region a homogeneous state is reached.
In particular, since the interfaces do not move there, the temperature becomes indeed constant.

For simplicity, we discuss here only the specific case of a purely diagonal eigenstrain, i.e.~$\epsilon_{ij}=0$ for $i\neq j$, to demonstrate the analytical procedure.
In contrast, in the case of an eigenstrain tensor with nonzero off-diagonal elements, bicrystal patterns emerge, and we will briefly discuss this situation below;
it is quite obvious that the analysis can be performed in an analogous way.

Due to the confinement of the system, phase coexistence is possible in a finite range of undercoolings, and we can calculate the asymptotic volume fraction $\fraction$ of the new phase (see Fig.~\ref{DilatationalSetup}).
Notice, that in contrast to the isothermal situation discussed in \cite{BrenerMarchenkoSpatschek2007}, the temperature in the tail region is not a control parameter but has to be found self-consistently, due to the thermally insulating boundaries.
In terms of the asymptotic phase fraction $\fraction$, the width of the $\beta$ phase in the tail region is written as $\fraction W$.
Mechanical compatibility requires
\begin{align}
(1-\fraction)\epsilon_{xx}^{[\alpha]}+\fraction\epsilon_{xx}^{[\beta]} & =0,\label{eq:MixedMode-coherency-cond_yy}
\end{align}
since the average lateral strain must vanish due to the boundary conditions for the displacement.
Together with the mechanical equilibrium condition of force balance, we can then calculate the elastic fields in the tail region, and consequently the elastic energy.
Then, we calculate the energy excess $\delta \cal{F}$, due to the appearance of a finite fraction of the new phase $\fraction$.
It is defined as the difference between the energy for $\fraction=0$ without any phase $\beta$ and the energy with an arbitrary but finite fraction,
\begin{align} \label{qwert}
\delta \cal{F}=&-W\left(\fraction F_{el}^{[\beta]}+(1-\fraction)F_{el}^{[\alpha]}\right)-2\gamma+W\int\limits _{0}^{\lambda}\left(\frac{L(T_{eq}-T(\lambda'))}{T_{eq}}\right)d\lambda'.
\end{align}
Initially, there is only the elastically relaxed but metastable phase $\alpha$ at the constant temperature $\Delta$.
As soon as we have a finite phase fraction $\fraction$, the energy increases by elastic and capillary contributions.
Finally, the integral appears due to thermal insulation, since an increase of the amount of the phase $\beta$ by phase transformation processes generates latent heat, and therefore causes an increase of the temperature. 
Using the heat conservation condition we find the relation between this temperature and the phase fraction to be
\begin{align}
T(\fraction)&=T_{\infty}+\fraction\frac{L}{\cheat}.
\end{align}
Therefore, we obtain for the energy excess as a function of the asymptotic phase fraction $\fraction$
\begin{align} \label{abc123}
\delta \cal{F}(\fraction)&=\frac{WL^{2}}{\cheat T_{eq}}\left((\Delta-\Delta_{el})\fraction -\frac{1}{2}\left(1+a_{el}\Delta_{el}\right)\fraction^{2}-\frac{2d_{0}}{W}\right),
\end{align}
where the parameter $\Delta_{el}$, as already introduced above in Eq.~(\ref{eq:Definition-DeltaEL}), defines the strength of elastic effects, and 
\begin{align}
a_{el}&=\frac{\cheat T_{eq}(1+\nu)(1-2\nu)}{L^{2}(1-\nu)E}\frac{\left(\sigma_{yy}^{0}\right)^{2}}{\Delta_{el}} \label{Free-Energy-Arb-Fraction}
\end{align}
is a parameter depending on the type of eigenstrain.
The maximum of this energy excess, $\partial \Delta F (\fraction)/\partial \fraction=0$, provides the equilibrium fraction, which expresses the thermodynamic balance between elastic deformation and free energy release due to the phase transition,
\begin{align}
\fraction_{eq}&= \frac{1}{1+a_{el}\Delta_{el}}\left(\Delta-\Delta_{el}\right). \label{eq:Fraction-of-beta-phase}
\end{align}
If the fraction $\fraction_{eq}$ is in the range $0<\fraction_{eq}<1$, we obtain phase coexistence in the asymptotic regime far behind the tip.
We would like to point out that one can derive this equation also by using the phase equilibrium condition Eq.~(\ref{eq:Phase-equilibrium-condition}).

Then growth of the $\beta$ phase demands that the energy excess (\ref{abc123}) for the equilibrium fraction (\ref{eq:Fraction-of-beta-phase}) has to be positive,
\begin{align}
\delta \cal{F}(\fraction)&=\frac{WL^{2}}{\cheat T_{eq}}\left(\frac{1}{2}\left(1+a_{el}\Delta_{el}\right)\fraction^{2}-\frac{2d_{0}}{W}\right) >0. \label{eq:Free-Energy-Gain}
\end{align}
This is equivalent to the condition $\fraction>\fraction_{crit}$, where $\lambda_{crit}$ is given by
\begin{align}
\fraction_{crit}^{2}&=\frac{4d_{0}}{W}\frac{1}{1+a_{el}\Delta_{el}}. \label{eq:Delta-Equilibrium}
\end{align}

We note that for the shear transformations discussed above, the most favorable process is the growth of a bicrystal, where two twinned phases $\beta$ and $\beta'$ with $\vartheta=\pm 2\pi/3$ grow together, because this is energetically most favorable \cite{BrenerMarchenkoSpatschek2007}.
For the bicrystal configuration, the analysis can be performed in a similar way, and then also the $\beta/\beta'$ grain boundary energy appears in the energy balance analogous to Eq.~(\ref{qwert}).


Considering the channel geometry, we use a phase field formulation to solve the full free boundary problem numerically.
This method is very versatile and allows to study even more complicated scenarios, e.g.~the full dynamical evolution and not only the steady state regime, three-dimensionsal situations with full anisotropy of elasticity and surface tension or simulations beyond the symmetrical model, where the two phases have different diffusion constants.
Nevertheless, for brevity of presentation, we confine our investigations to the simple, isotropic, two-dimensional and thermally symmetric case that we also used for the Green's function approach in the free space geometry and the above analytical calculations.

The phase field method introduces an additional numerical lengthscale $\xi$, the width of the transition region between the different solid phases.
This means that an order parameter, which discriminates between the different phases, has a profile $\phi=[1+\tanh(n/\xi)]/2$ in normal direction through an interface.
We operate here in the sharp interface limit, which means that this lengthscale is significantly smaller than all other physical lengthscales present in the problem.
Then the description recovers the desired sharp interface equations that were proposed above.

Hence, we introduce the phase field $\phi$ having a value $\phi=1$ in the metastable initial phase, and a value $\phi=0$ in the new phase.
We start from a free energy functional
\begin{align}
F[\phi,w,u_{i}] & =\int\limits _{V}\left(f_{s}+f_{dw}+f\right)dV,\label{eq:PhaseFieldModel-free-energy-functional}
\end{align}
where $f_{s}(\nabla\phi)=3\gamma\xi(\nabla\phi)^{2}/2$ is the gradient energy density and $f_{dw}(\phi)=6\gamma\phi^{2}(1-\phi)^{2}/\xi$ is the double well potential, guaranteeing that the free energy functional has two local minima at $\phi=0$ and $\phi=1$ corresponding to the two distinct phases of the system.
The form of the double well potential and the gradient energy density are chosen such that the phase field parameter $\xi$ defines the interface width and the parameter $\gamma$ 
corresponds to the interface energy of the sharp interface description \cite{Gugenberger08}. 
The free energy density $f(\phi,w,u_{i})$ of our phase field description is chosen to be an interpolation between the two distinct free energies of the model,
\begin{align}
f(\phi,u_{i},w)= & F^{[\alpha]}_{el}h(\phi)+F^{[\beta]}_{el}(1-h(\phi)) + L\frac{T-T_{eq}}{T_{eq}} (1-h(\phi)),\label{eq:PhaseFieldModel-free-energy-interpolation}
\end{align}
where we choose the interpolation function $h$ to be $h(\phi)=\phi^{2}(3-2\phi)$. It is the simplest polynom satisfying the necessary interpolation conditions $h(0)=0$ and $h(1)=1$, and having a vanishing slope at $\phi=0$ and $\phi=1$, in order not to shift the bulk states. 

The evolution equation of the phase field is given by the variational expression
\begin{align}
\frac{\partial\phi}{\partial t} & =-\frac{M}{3\gamma\xi}\left(\frac{\delta F}{\delta\phi}\right)_{u_{i},w}.
\end{align}
For large values of the mobility $M$ we recover the desired case of diffusion limited growth.
The elastodynamic equations read
\begin{align}
\rho\ddot{u}_{i} & =- \left(\frac{\delta F}{\delta u_i}\right)_{\phi, w} = \frac{\partial\sigma_{ik}}{\partial x_{k}},
\label{eq:Intro-Phasefield-elastodynamic-equations}
\end{align}
which recover static elasticity for slowly propagating interfaces in comparison to the sound speed $v_s\sim (E/\rho)^{1/2}$, with $\rho$ being the mass density.

For the temperature field we have the usual thermal diffusion equation with the motion of the phase field or interface being a source of latent heat,
\begin{align}
\frac{\partial w}{\partial t} & =D\nabla^{2}w+h'(\phi)\frac{\partial\phi}{\partial t},
\end{align}
where the prime denotes the derivative with respect to $\phi$.
The phase field model presented here is very similar to the model in \cite{Slutsker2006}. 
However, there only the very special case of melting or crystallization of a confined sphere with elastic effects was discussed.

For the simulations we typically use parameters $W/d_0=100$ and $W/\xi=80$.

\section{Results}
\label{results}

Apart from the theoretical investigations above, we also performed numerical simulations for both the free growth in an infinite system and in a channel or narrow strip.
As already mentioned before, we observe the growth of a single new phase if the eigenstrain tensor is purely diagonal, and twin structures for shear transformations with $\vartheta=\pm 2\pi/3$.
Also, we consider mixed mode situations which are between these two extreme cases.

\subsection{Single crystal}

We first consider the situation of free growth.
Here, the simplest case is that of a dilatational eigenstrain where all diagonal elements have the same value, i.e.~$\epsilon_{ik}^{0}=\epsilon\delta_{ik}$; 
this case is obtained from Eq.~(\ref{MixedModeEigenstrainDefinition}) by setting $\eta=1$. 
As already mentioned before, selection of a steady solution in the absence of elastic effects is not possible for isotropic surface tension.
It is quite remarkable that also the inclusion of elastic effects due to the dilatational eigenstrain of the transition does not change this situation;
the reason is that the elastic effects can be mapped to a change of the undercooling only, and therefore we recover again the missing selection of conventional free dendritic growth \cite{DenisBrenerMe2008}.
It turns out that also for the volume-preserving shear strain transitions with $\epsilon_{yy}^0=-\epsilon_{xx}^0=\epsilon$, where all other components $\epsilon_{ik}^0$ vanish, i.e. $\eta=0$ and $\vartheta=0$, steady state solution do not exist either.

\begin{figure}
\begin{centering}
\includegraphics[angle=0,width=0.45\textwidth]{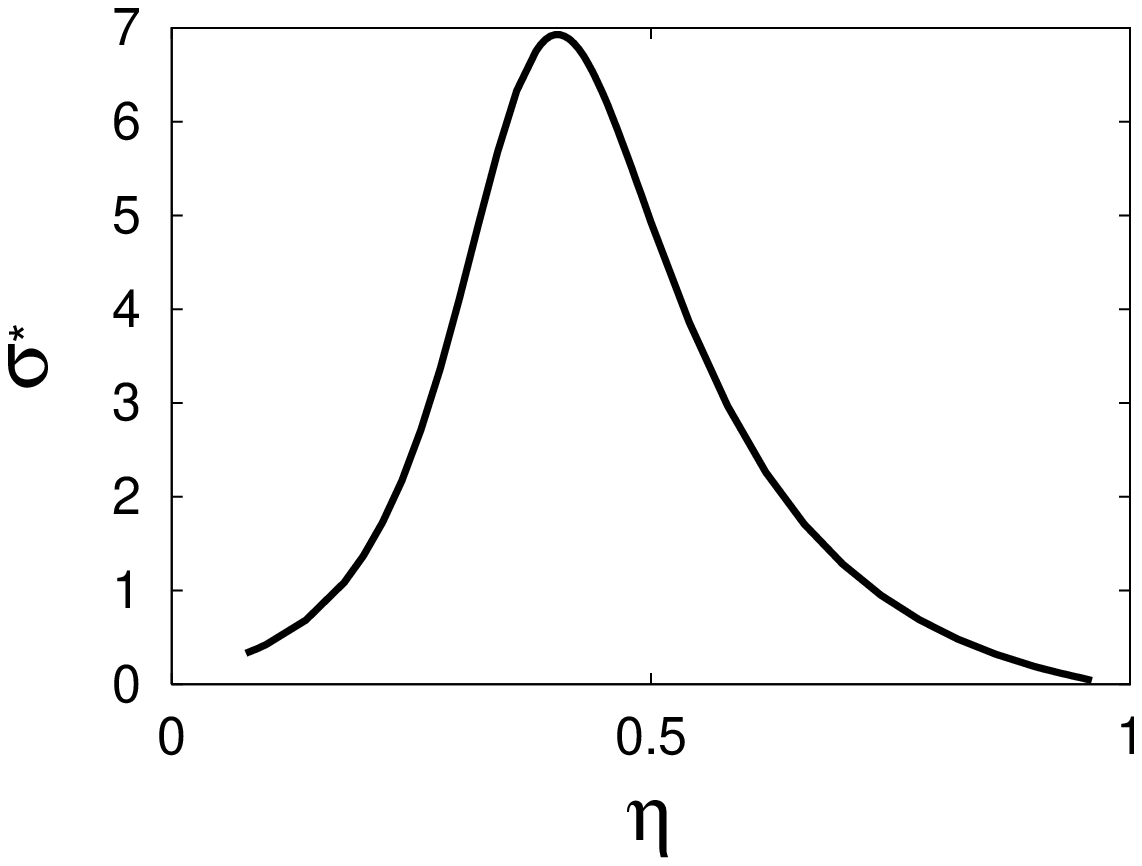}
\includegraphics[angle=0,width=0.45\textwidth]{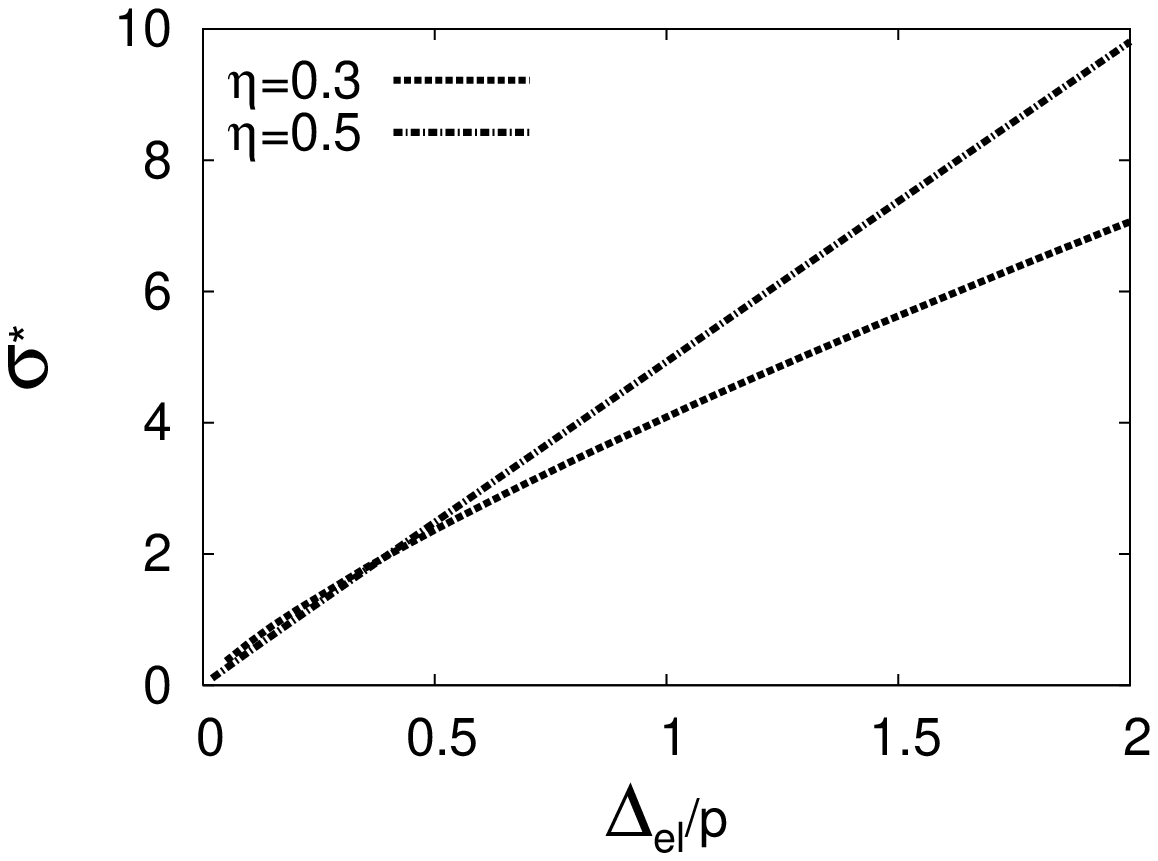}
\end{centering}
\caption{The left panel shows the stability parameter $\sigma^{\ast}$ (see Eq.~(\ref{eq:stabilityParameterDefinition})) as function of the mixing parameter $\eta$ for the privileged growth direction with an eigenstrain tensor $\epsilon_{ik}^{-}$ when $\Delta_{el}/p=1$.
On the right panel we show the stability parameter $\sigma^{\ast}(\Delta_{el}/p)$ for two different values of mixing $\eta$.}
\label{MixedModeResultSingleCrystal2} 
\end{figure}

Under these circumstances it is very suprising, that for values of the mixing parameter in between zero and one $0<\eta<1$, $\vartheta=0$  we do find steady state solutions for free growth and negative mixing. 
This is shown in Fig.~\ref{MixedModeResultSingleCrystal2}a, where the selected ``stability parameter'' $\sigma=d_0/pR=\sigma^{\ast}$, which is a dimensionless measure for the growth velocity, is plotted as function of the mixing parameter $\eta$.
In general, the eigenvalue $\sigma^{\ast}$ depends on $\eta,\Delta_{el}$ and $p$, but in the regime of small Peclet numbers the elastic driving force and $p$ appear only in the combination $\Delta_{el}/p$.
Fig.~\ref{MixedModeResultSingleCrystal2}b shows the dependence of the selected stability parameter on $\Delta_{el}/p$ for two different values of $\eta$.
For the other type of mixing $\epsilon_{ik}^{+}$ corresponding to the unprivileged direction of growth, we do not find steady state solutions in free space.
We note that from the selected stability parameter $\sigma^*$ one can extract the steady state growth velocity $v$ via the relation
\begin{equation} \label{eq:stabilityParameterDefinition}
v=\frac{2D}{d_0} \sigma^* p^2.
\end{equation}

\begin{figure}
\includegraphics[angle=0,width=0.45\textwidth]{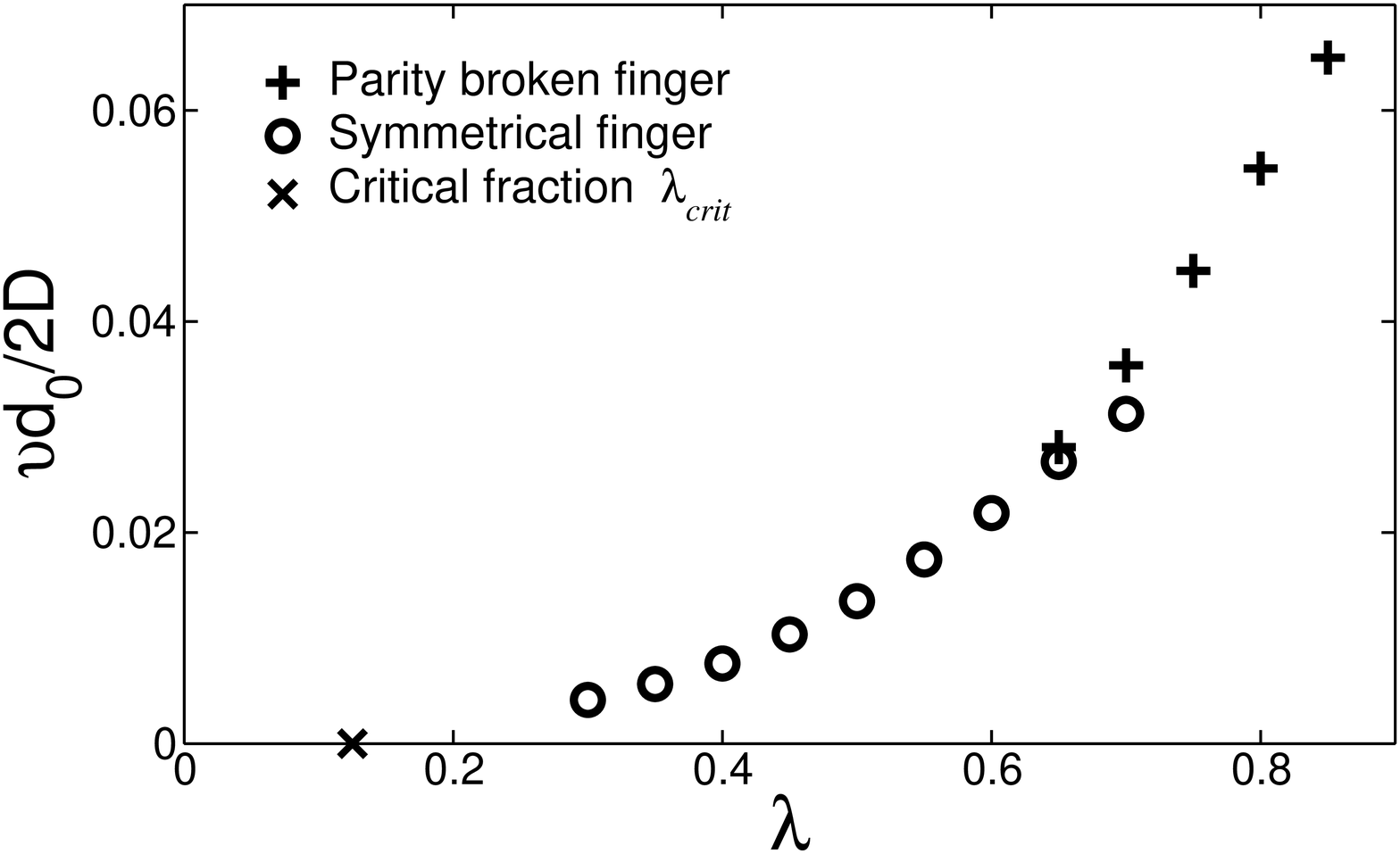}
\includegraphics[angle=0,width=0.45\textwidth]{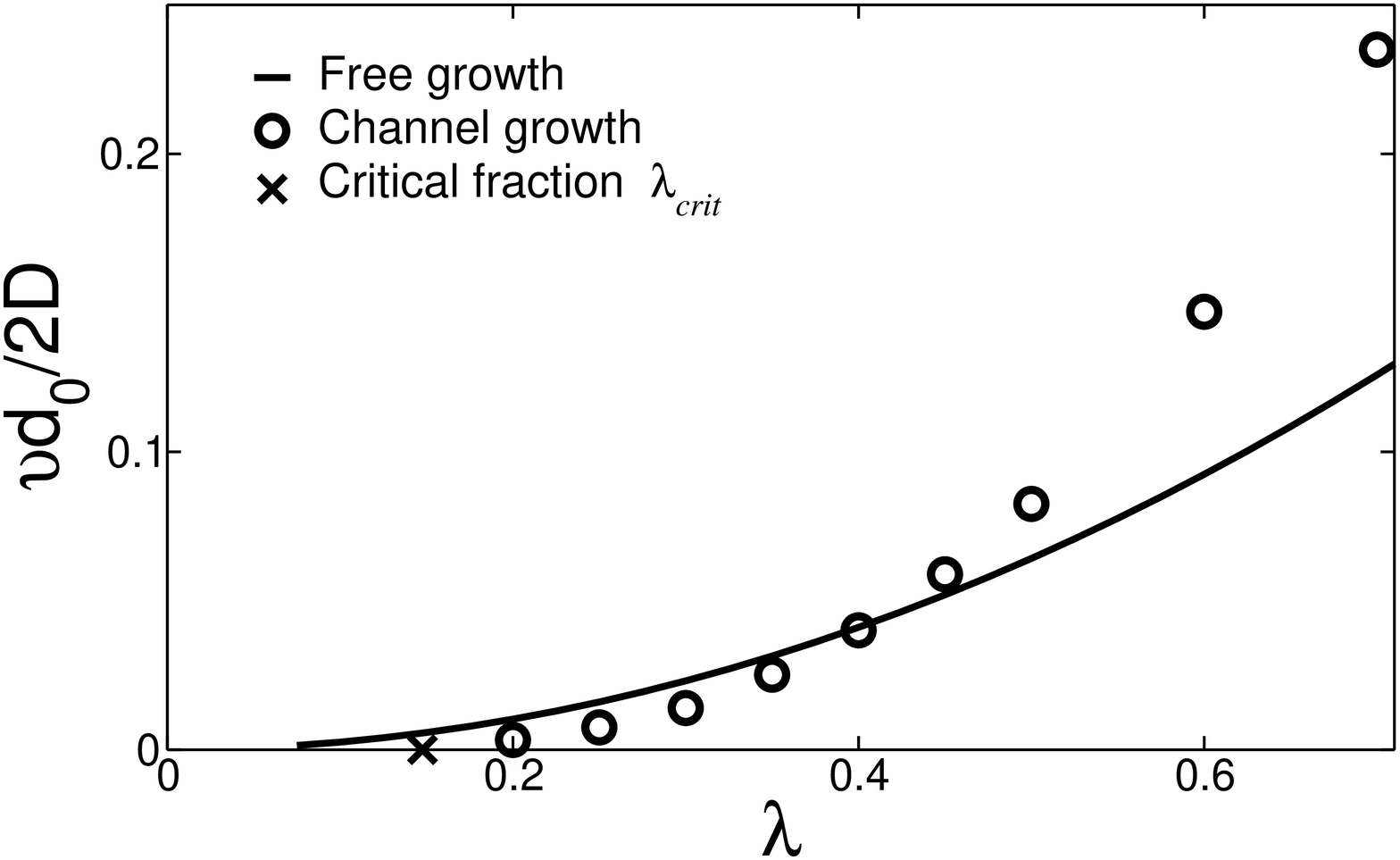}
\caption{Channel growth with diagonal eigenstrain. 
The left panel shows the dependence of the growth velocity on the driving force $\fraction$, where we choose a dilatational eigenstrain $(\eta=1, \Delta_{el}=0.4)$. 
For small driving forces we obtain velocities and shapes corresponding to a symmetric solution (see Fig.~\ref{DilatationalSetup}$a$). 
For higher driving forces a first order transition towards asymmetrical solutions takes place (see parity broken finger morphology in Fig.~\ref{DilatationalSetup}$b$).
In the right panel the comparison of growth velocities between phase field and the free space calculations is shown, for the case of negative mixing corresponding to the privileged growth direction $(\eta=0.5, \vartheta=0, \Delta_{el}=0.05)$. 
The agreement is reasonably good, especially for driving forces $\fraction$ slightly above the critical phase fraction $\fraction_{crit}$ of zero growth velocity.}
\label{fig:DiagonalEigenstrain}
\end{figure}

In contrast, for the growth in a channel, steady state solutions do also exist for the case of a dilatational eigenstrain, $\eta = 1$ (see Fig.~\ref{fig:DiagonalEigenstrain}a). 
As expressed through Eqs.~(\ref{eq:Free-Energy-Gain}) and (\ref{eq:Delta-Equilibrium}), the phase $\beta$ can not grow for phase fractions $\fraction$ below the threshold $\fraction_{crit}$.
Beyond this value the growth of a symmetrical finger (dendrite) is the most favorable solution.
We note that in situations where free growth solutions do not exist, the still existing solutions in the channel must have a scale that is determined by the size $W$ of the channel.
For higher driving forces an asymmetric finger grows with higher velocities, and is therefore more likely to be observed;
typical finger shapes are shown in Fig.~\ref{DilatationalSetup}.

For the situation of $\eta = 0.5$, where steady state solutions exist also in the free space, we performed a quantitative comparison of the results of the two methods in Fig.~\ref{fig:DiagonalEigenstrain}b.
For this comparison we directly relate the fraction $\lambda$ to the driving force $\tilde{\Delta}$ for free growth via Eq.~(\ref{eq:Fraction-of-beta-phase}) and Eq.~(\ref{eq:Shifted-undercooling}).
Then, in the limit of small Peclet numbers, the latter is related to $\tilde{\Delta}$ according to $p \approx \tilde{\Delta}^2/\pi$.
Finally, via the calculated function $\sigma^{\ast}(\Delta_{el}/p)$ and Eq.~(\ref{eq:stabilityParameterDefinition}) we obtain the growth velocity as a function of $\fraction$, which is then compared to the phase field results.

The results for the growth velocity agree well in a range of intermediate fractions of the new phase $\beta$.
In particular, we confirmed that the steady state solutions which were obtained by the Green's function method are indeed stable, as the dynamical phase field method tracks only these stable branches.
For low phase fractions, growth is not possible below the threshold $\fraction_{crit}$, which tends to zero for infinitely wide channels.
For very high driving forces, the growing finger is influenced by the geometrical confinement of the channel width and also affected by kinetic corrections to the sharp interface equations of motion, which therefore lead to a deviation from the free growth result which was obtained by the Green's function method.

It is interesting to compare the situation to a case without elasticity, where the problem reduces to classical growth in a channel \cite{BrenerHMKSaitoTemkin, KesslerBenJakob1995, BenAmarBrener1995, BenAmarBrener1996, KesslerKoplikLevine1986, Geilikman}.
It is known that there even in wide channels symmetric solutions do not exist below a fraction $\fraction=1/2$, whereas we found that the eigenstrain leads to the existence of solutions at already significantly lower driving forces $\fraction$, see Fig.~\ref{fig:DiagonalEigenstrain}a.

\subsection{Bicrystal growth}

\begin{figure}
\begin{center}
\includegraphics[width=0.65\textwidth]{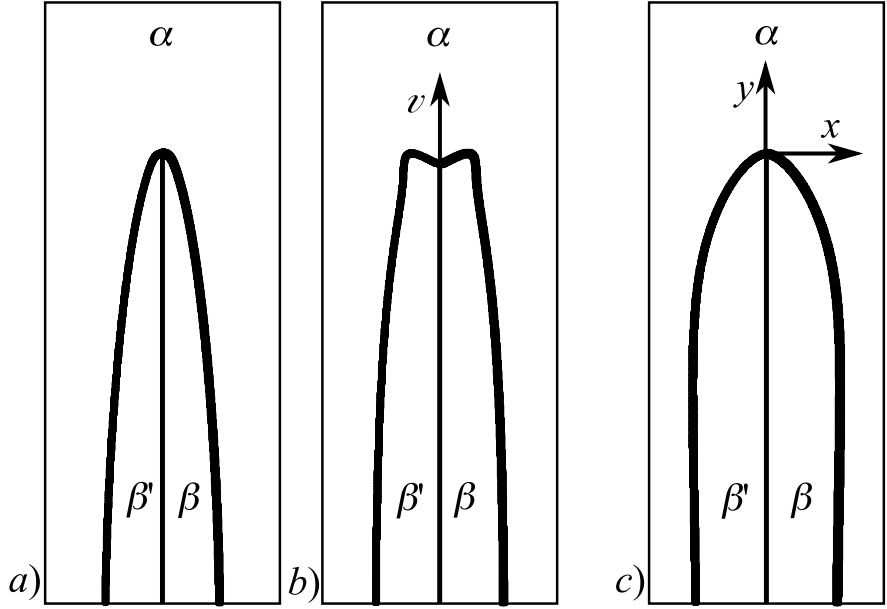}
\caption{Bicrystal patterns growing in a finite channel. 
$a)$ shows the single tip solution and $b)$ the twinned finger morphology, which corresponds to the most favorable solution branch for higher driving forces. 
The positive mixing chosen for $a)$ and $b)$ leads to growth along the privileged direction and the parameters are $\eta = 0.1, \Delta_{el}=0.1, \fraction=0.6$.
The panel $c)$ shows the shape of the slow growth direction for $\eta = 0.5, \Delta_{el}=0.2, \fraction=0.6$.
\label{BiCrystalShapes}}
\end{center}
\end{figure}

In contrast to the cases studied above, the presence of off-diagonal eigenstrain entries lead to the preferential growth of bicrystal patterns.
This behavior can easily be understood intuitively, since the eigenstrain in the $\beta$ phase favors the presence of off-diagonal strain. 
Then the boundary conditions of fixed displacements require that the surrounding $\alpha$ phase is sheared in the opposite direction.
This implies a finite elastic energy which is stored far behind the tip.
If, on the other hand, a second finger $\beta'$ of the same size with an off-diagonal eigenstrain with opposite sign appears, the elastic energy contribution that is associated with the off-diagonal elements of the strain tensor vanishes.
Consequently, the total elastic energy in the tail region is reduced, and the growth of the bicrystal therefore thermodynamically favored.

An additional energy contribution emerges from this bicrystal configuration, since the boundary between $\beta$ and $\beta'$ generates a grain boundary energy $\gamma_{gb}$.
We consider in particular the case of vanishingly small twin boundary energy, $\gamma_{gb} \ll \gamma$.
This affects also the geometrical situation at the tip, which is now a triple junction, and the contact angle $\phi$ becomes zero, so the overall finger has a smooth contour.
Typical shapes of such bicrystal pattern are shown in Fig.~\ref{BiCrystalShapes}, which are here related to the growth in a channel, but they look similar for free growth.
We note that we simulated only half of the channel using the symmetry of the growing structures \cite{BrenerMarchenkoSpatschek2007}.

\begin{figure}
\begin{center}
\includegraphics[angle=0,width=0.6\textwidth]{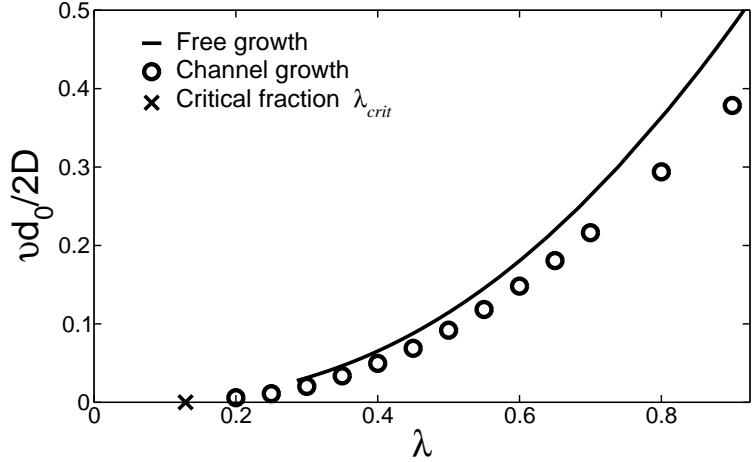}
\caption{Bicrystal with pure shear eigenstrain, $\eta = 0$: 
Comparison of growth velocities obtained from the boundary integral technique for the free space geometry \cite{DenisBrenerMe2008} with the velocities from phase field simulations for the channel geometry ($\Delta_{el}=0.3$).
\label{fig:BicrystalPureShearComp}}
\end{center}
\end{figure}

Considering the free space geometry, the first important observation is that in contrast to the single crystal scenario, steady state solution do exist also for pure shear, $\eta=0$, which is already discussed in \cite{DenisBrenerMe2008}.
New is here that we found the same morphology also in the channel geometry, showing surprisingly good agreement in the comparison of growth velocities, done in the same way as was done for the single crystal described above (see Fig.~\ref{fig:BicrystalPureShearComp}).
Again, we would like to point out that the stability of the bicrystal solution obtained from the Green's function technique is confirmed by the phase field simulations.

\begin{figure}
\begin{centering}
\includegraphics[angle=0,width=0.6\textwidth]{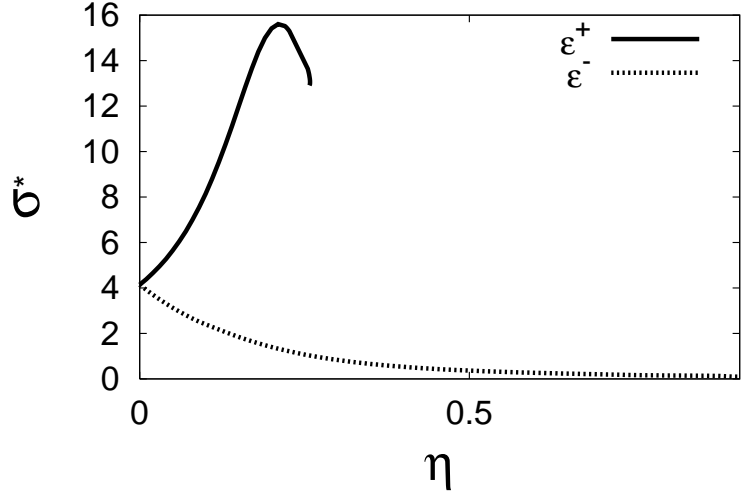}
\par\end{centering}
\caption{Plot of the stability parameter $\sigma^{\ast}(\eta)$ as function of the admixture parameter $\eta$, for the two different types of mixing and $\Delta_{el}/p=1.0$.
The solid line corresponds to the privileged direction of growth, i.e.~positive mixing, while the dotted line shows the dependence for the unprivileged negative mixing configuration.}
\label{BiCrystalSigmaOfEta} 
\end{figure}

\begin{figure}
\begin{centering}
\includegraphics[angle=0,width=0.45\textwidth]{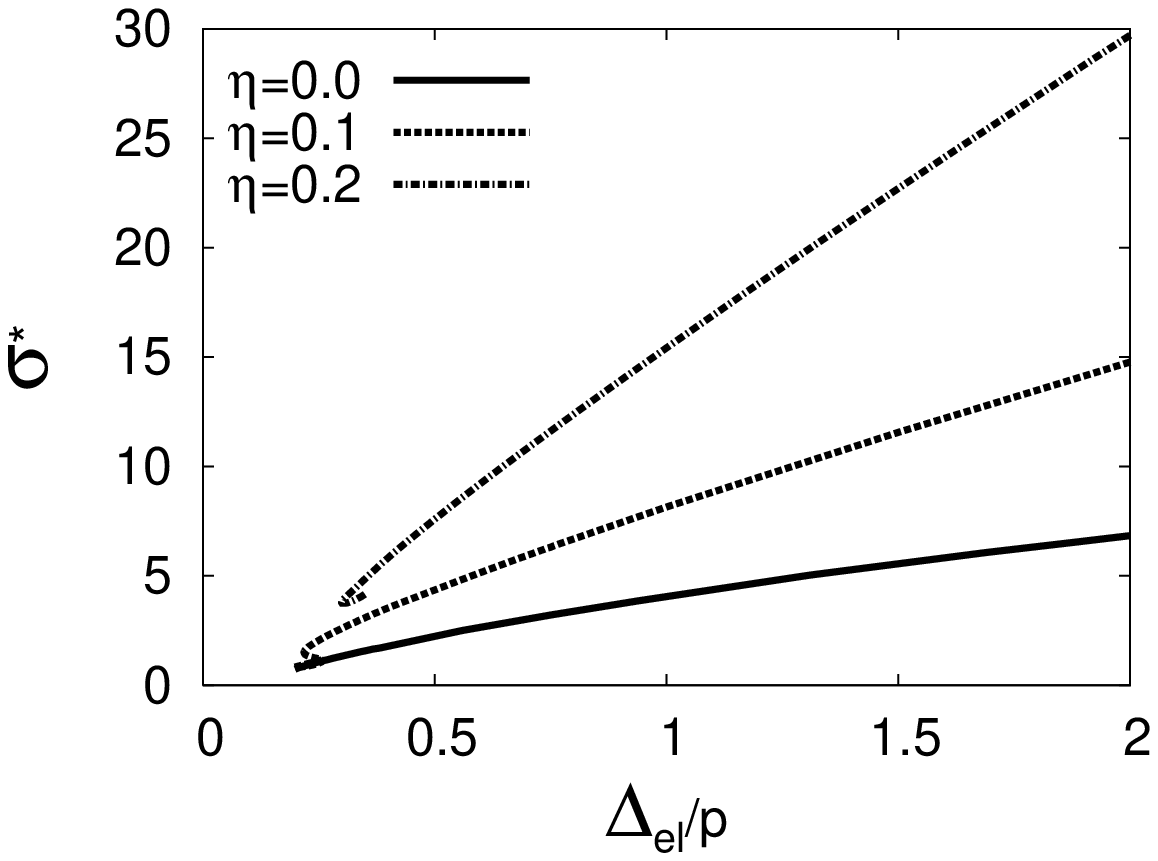}
\includegraphics[angle=0,width=0.45\textwidth]{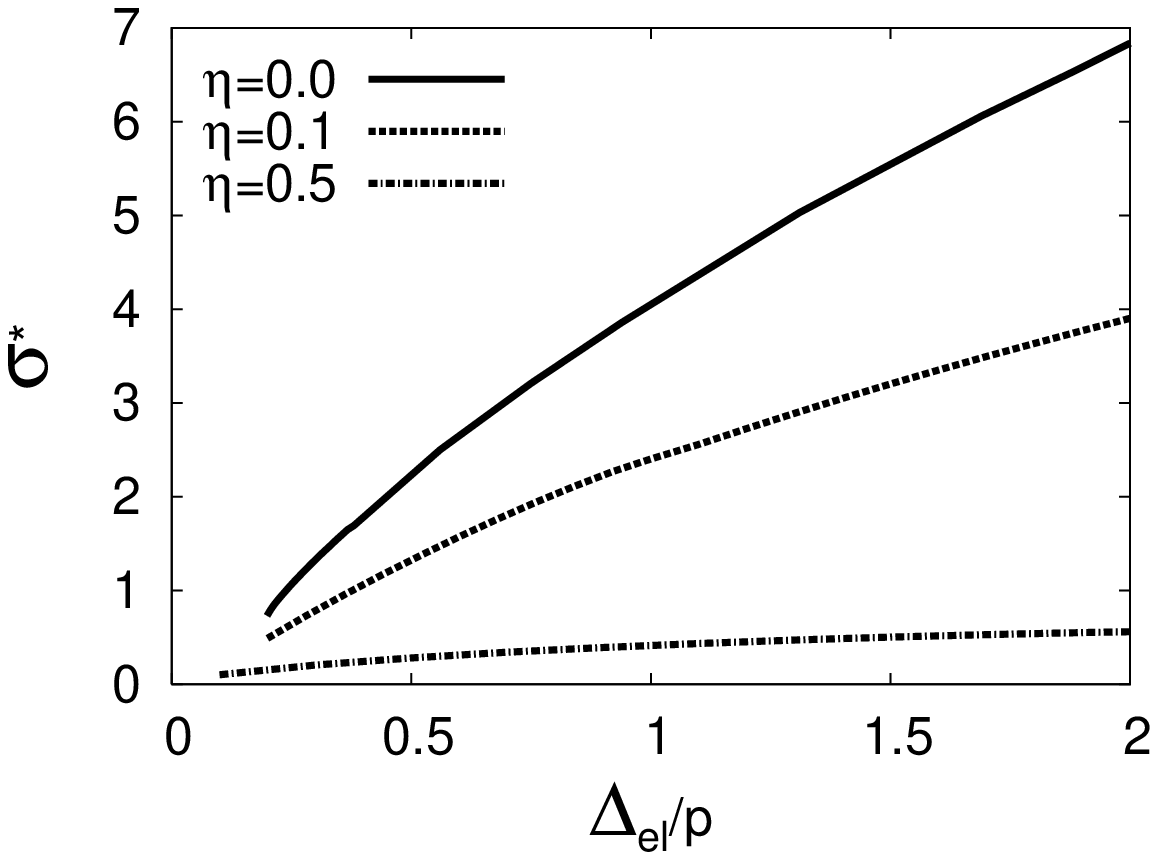}
\par\end{centering}
\caption{The left panel shows the stability parameter $\sigma^{\ast}$ as function of the control parameter $\Delta_{el}/p$, for positive mixing, whereas the in the right panel $\sigma^{\ast}(\Delta_{el}/p)$ is plotted for negative mixing. Additionally, both graphs contain the $\sigma^{\ast}(\Delta_{el}/p)$ dependence for the pure shear configuration $\eta=0$, as a reference to compare both types of mixings, which was taken from \cite{DenisBrenerMe2008}. 
}
\label{BiCrystalPositiveAndNegativeDilatInShearSigmaDeltaEl} 
\end{figure}

We also did investigations of the bicrystal free growth considering mixed eigenstrains, and the results are summarized in Fig.~\ref{BiCrystalSigmaOfEta} and \ref{BiCrystalPositiveAndNegativeDilatInShearSigmaDeltaEl}.
In the first graph, the stability parameter $\sigma^*$, which is again a measure for the dimensionless velocity, is shown as function of the admixture parameter $\eta$ for fixed ratio $\Delta_{el}/p$.
As before, the elastic driving force and the $p$ appear only the combination $\Delta_{el}/p$ for small Peclet numbers.


As already mentioned above it is quite interesting that the appearance of an additional dilatational strain leads to an effective anisotropy of the system:
Whereas in the case of a positive admixture $\epsilon^{+}$ the velocity grows with increasing $\eta$, the opposite happens for negative admixture $\epsilon^{-}$.
This suggests that in free growth the bicrystal structure will eventually grow in the direction with the higher propagation velocity.
For the fast branch, we did not find solutions beyond the point $\eta\approx 0.25$.

\begin{figure}

\includegraphics[angle=0,width=0.45\textwidth]{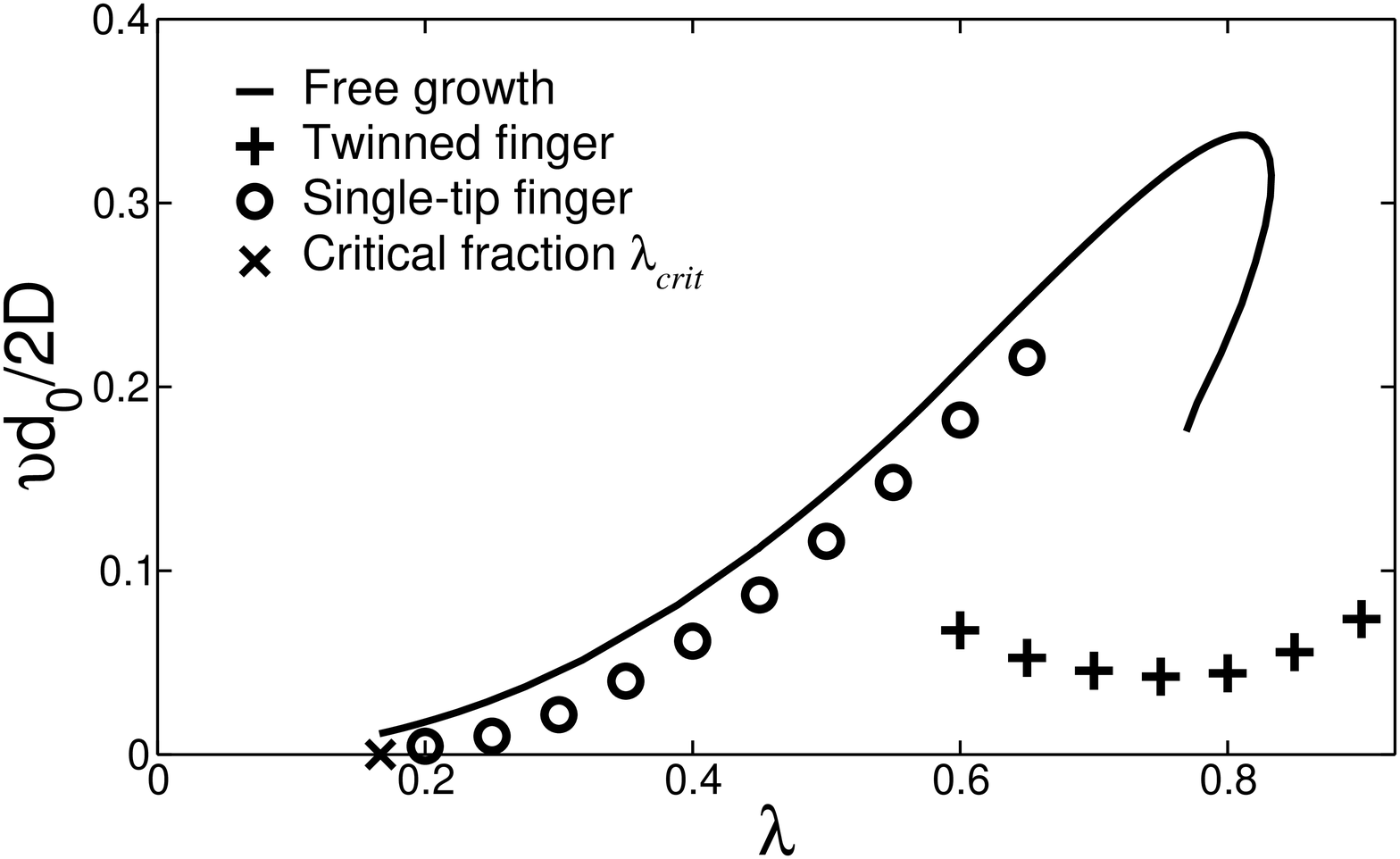}
\includegraphics[angle=0,width=0.45\textwidth]{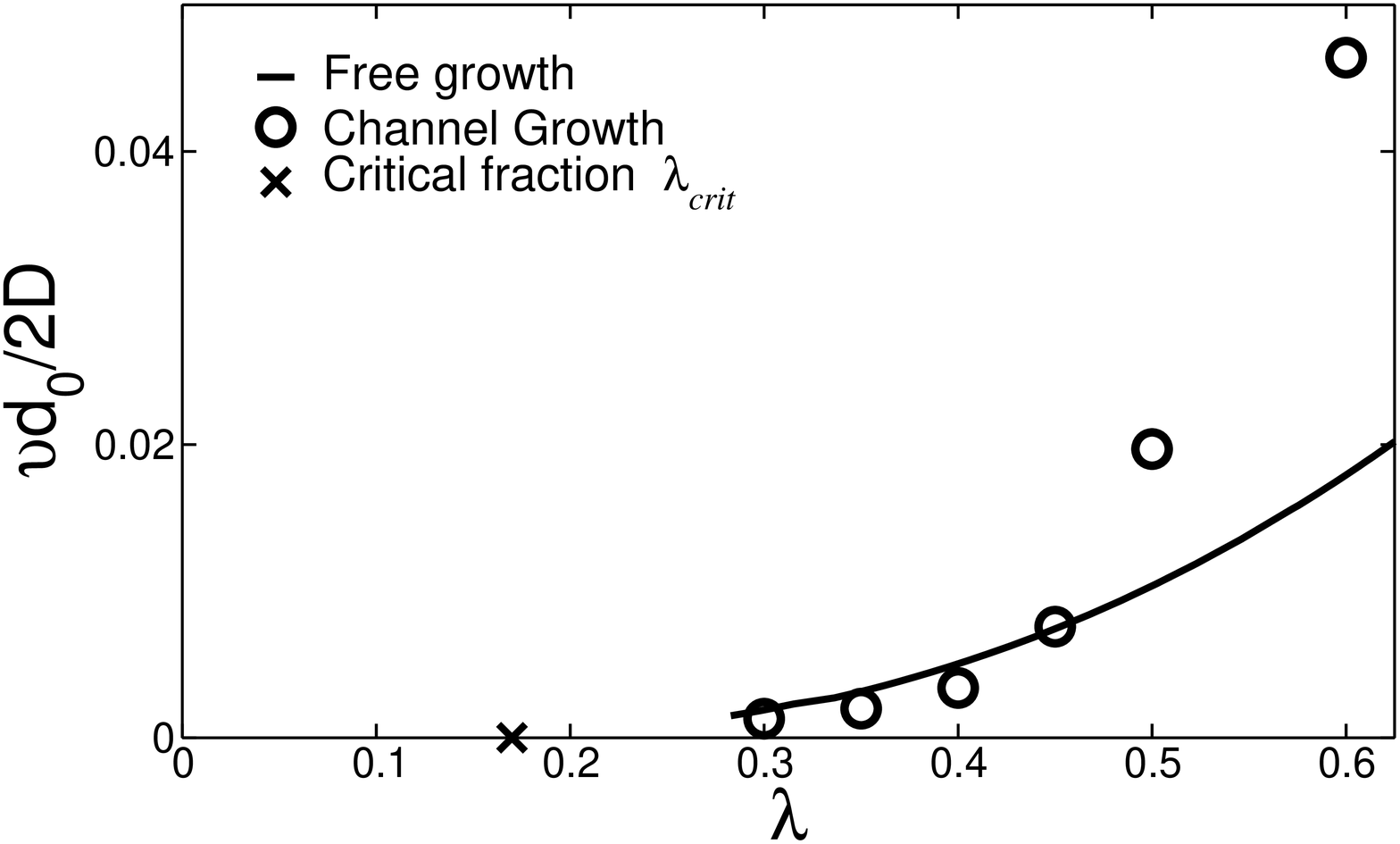}

\caption{Plot of the dimensionless growth velocity as function of the ``driving force'' $\lambda$ for bicrystal growth with mixed eigenstrain: 
comparison of free growth results obtained by Green's function methods to phase field simulations for the growth in a channel. 
Left panel: positive admixture with $\eta = 0.1$. Right panel: negative admixture, $\eta = 0.5$.}
\label{fig:BiCrystallMixedModeCompChannelAndFreeGrowth}
\end{figure}

In Fig.~\ref{fig:BiCrystallMixedModeCompChannelAndFreeGrowth} we show the comparison of the two geometries for the bicrystal case.
Again, the phase field results confirm the stability of the solution branch for low driving forces.
In this region, we obtain a convincing quantitative agreement between the growth velocities obtained for the free growth scenario using the Green's function method, and the channel geometry with the phase field method.
In the left panel the favored direction is shown (positive admixture) for $\eta=0.1$.
In this case we again observe dendritic solutions for small driving forces  $\fraction$ (see also Fig. \ref{BiCrystalShapes}a).
For the chosen parameters in the channel geometry, all solutions on the upper branch are dendritic.
On the lower branch, however, the steady state shapes are double-fingers, which means here the curvature directly at the point of symmetry has changed its sign and we obtain two tips (see Fig. \ref{BiCrystalShapes}b).
The velocity of these branched solutions is here always smaller than for the single-tip structures.

For the other direction, i.e. negative admixture, and a mixing parameter $\eta=0.5$, we show the comparison of growth velocities in the right panel of Fig.~\ref{fig:BiCrystallMixedModeCompChannelAndFreeGrowth}.
Here, we obtain a convincing agreement between the different methods and associated geometries at least for small driving forces, indicating dynamical stability also for this type of solution.
Obviously, the growth velocities are here substantially lower than for the previous case, and for fully free growth an initial bicrystal seed would therefore preferentially grow in the ``fast" direction.

\section{Summary and Conclusion}

In summary, we investigated the diffusion-limited kinetics of solid-solid transformations in the presence of lattice strain.
We studied in particular the free growth with Green's function methods and the growth in a channel by phase field techniques, exhibiting a comparable behavior.
It turns out that the elastic effects have a strong influence on the selection of the arising patterns and the corresponding steady state velocities, which differ crucially from conventional dendritic growth.
In particular, we find that the selected stability parameter and the corresponding growth velocity can be much higher compared to the classical case of dendritic growth, where the solutions are selected by the tiny effect of the anisotropy of surface tension.
Different structural transitions, which are described by dilatational and shear eigenstrains, lead to a rich behavior, and we find very different structures already for the simple cases that were discussed here.

Here, we focused on growth situations that are controlled by thermal diffusion, but it is known that the also interesting scenario of chemical diffusion can be mapped to this case.

In comparison to the Green's function method, which is computationally very efficient, the phase field technique has the advantage of higher flexibility.
In particular, it can easily be extended to three-dimensional cases, it can be used also for non-steady state situations or anisotropic or non-symmetrical situations with different material parameters in the phases.

This work was supported by the DFG grant SPP 1296 and the German-Israeli Foundation.

\end{document}